# On the Fibril Elongation Mechanism of the Prion Protein Fragment PrP106-126

Xin Zhao, Shuo-Xing Dou, Ping Xie, and Peng-Ye Wang[*]
Laboratory of Soft Matter Physics, Beijing National Laboratory for Condensed Matter Physics, Institute of Physics, Chinese Academy of Sciences, Beijing 100080, China

Abstract

Mouse prion protein PrP106-126 is a peptide corresponding to the residues 107-127 of human prion protein. It has been shown that PrP106-126 can reproduce the main neuropathological features of prionrelated transmissible spongiform encephalopathies and can form amyloid-like fibrils in vitro. The conformational characteristics of PrP106-126 fibril have been investigated by electron microscopy, CD spectroscopy, NMR and molecular dynamics simulations. Recent researches have found out that PrP106-126 in water assumes a stable structure consisting of two parallel β-sheets that are tightly packed against each other. In this work we perform molecular dynamics simulation to reveal the elongation mechanism of PrP106-126 fibril. Influenced by the edge strands of the fibril which already adopt β-sheets conformation, single PrP106-126 peptide forms β-structure and becomes a new element of the fibril. Under acidic condition, single PrP106-126 peptide adopts a much larger variety of conformations than it does under neural condition, which makes a peptide easier to be influenced by the edge strands of the fibril. However, acidic condition dose not largely affect the stability of PrP106-126 peptide fibril. Thus, the speed of fibril elongation can be dramatically increased by lowering the pH value of the solution. The pH value was adjusted by either changing the protonation state of the residues or adding hydronium ions (acidic solution) or hydroxyl ions (alkaline solution). The differences between these two approaches are analyzed here.

## 1 Introduction

The cellular prion protein ($PrP^C$) is a neuronal cell-surface glycoprotein. It is bound to the plasma membrane by glycosylphosphatidylinositol anchor [1]. A conformationally altered form of $PrP^C$, called the scrapie form prion protein ($PrP^{Sc}$), can cause various transmissible spongiform encephalopathies (TSEs), such as bovine spongiform encephalophathy in cattle and scrapie in sheep [2,3]. $PrP^{Sc}$ can aggregate and form amyloid plaques in the brain. The structure of $PrP^{Sc}$ has been investigated by NMR experiments [4-8], whereas structures of $PrP^{Sc}$ remain elusive. The two isomers of PrP have dramatically different structure: $PrP^C$ is almost helical, with 47% α-helix and 3% β-structure. However, $PrP^{Sc}$ contains about 40% β-structure [9,10] and is proteinase K resistant (PrP-res) [9,11,12].

---

Prion protein is composed of a structured globular C-terminal domain and an unstructured N-terminal. Analysis of deletion mutants showed that only deletion of segments 95-107, 108-121 or 122-140 could abolish the conformational transition from $PrP^C$ to $PrP^{Sc}$ [17]. The fragment 106-126 is located within this region and has been found to adopt different conformations under different solution conditions [18,19]. This fragment of prion protein can reproduce the main neuropathological features of prion-related transmissible spongiform encephalopathies [20,21] and thus is used to investigate the mechanism of fibril formation and prion induced cell death. Experimental investigations found that PrP 106-126 and similar fragments can form fibril under acidic solution conditions [14,22,23]. (Researches on PrP also showed that acidic pH is capable of inducing the conformational change from $PrP^C$ to $PrP^{Sc}$ [13,14] and molecular dynamics simulations [15,16] are consistent with these results.) Electron Microscopy, CD Spectroscopy, NMR Spectroscopy, as well as molecular dynamics simulations have been used to identify the structural properties of the fragment 106-126 fibril [23-25]. It is found out that the fibril consists of two parallel β-sheets that are tightly packed against each other [23,25]. However the mechanism of fibril elongation and how acidic solution condition influences the formation of the fibril remain unanswered. In this work, we try to uncover how acidic condition speed up the elongation of the fibril by molecular dynamics simulation.

In previous works concerning molecular dynamics simulation of proteins, the effect of pH is taken into consideration by changing the "protonation state" of the proteins [15], i.e. the ionization situation of the side-chains. However, we control the pH value by adding hydronium ions or hydroxyl ions, for the following three reasons:

1. In this polypeptide fragment, only two of the 21 residues can have different protonation state under different pH values. Changing the protonation state of these two residues does have influence on the behavior of the peptide. But in this case, the pH value is not easy to control. The R group of His residue has a pK = 6.00. Since our system included only 1-3 peptides, the difference between some pH values (for example, pH=2 and pH=4, which means 1/100 and 1/10000 His residues are differently ionized respectively) can not be distinguished by changing the protonation state of the residues.

2. The properties of this peptide are mainly determined by the hydrophobic residues which do not have different protonation states.

3. The required conditions to observe the effects of pH values within the limited simulation time are actually pH=1-2 and pH=13-14.

By adding hydronium ions or hydroxyl ions we demonstrate how these two ions affect the structure of water and fragment 106-126. When surrounded by pure water, the fragment adopts a hairpin structure in most of the simulation time and this prevent the elongation of the fibril. When we add hydronium ions into water, the fragment adopts a more variety of structures, which facilitate the influence of the edge strands of the fibril to this peptide. It then forms β-structure along the edge strands and becomes a new element of the fibril.

## 2 Simulation Methods

### 2.1 Peptide, Condition and Forces

The PrP 106-126 fragment sequence Thr-Asn-Met-Lys-His-Met-Ala-Gly-Ala-Ala-Ala-Ala-Gly-Ala-Val-Val-Gly-Gly-Leu-Gly-Gly is mainly composed of hydrophobic residues and is found to assume a double β-sheet fibril in water in previous investigations [23,25]. In our simulation, all the atoms within the peptide and water solution are considered explicitly, and the interactions are computed using the GROMOS-96 force field [27,33]. Bond stretching, bending angles, dihedral potentials and non-bonded (van der Waals and electrostatic) interactions with 1nm-cutoff are included. We used SPC model for water molecules [28]. The MD simulations were performed in the canonical NPT (number of particles-pressure-temperature) ensemble at 300K with periodical boundary condition using software GROMACS [26], Version 3.2.1. Typical system size is 200 $nm^3$ with 5000-10000 water molecules and density 1.00 $g/cm^3$. The time step in the MD simulations is 2 fs.

## 2.2 Molecular Dynamics Simulation

### 2.2.1 Simulation of a Single Peptide

The starting structure of PrP 106-126 fragment for simulation is β-strand generated by software Hyperchem [29]. Since β-strand is not the peptide's stable structure in water, it takes about 400 ps for the system to reach equilibrium. The simulations are conducted for 6-8 ns, and the data from 2 ns to 6 ns is saved for analysis.

Simulations under neural, acidic and alkaline conditions are all conducted. The pH value is adjusted either by (1) changing the protonation state of Lys and His or by (2) changing the protonation state of Lys and His and adding hydronium ions or hydroxyl ions into the solution. The charge distribution and bond length of the hydronium ions and hydroxyl ions are obtained by *ab initio* quantum mechanics calculation using software Hyperchem [29]. In many times of the simulations, these two methods give consistent results, except for that the very low and very high pH values can not be achieved by (1).

### 2.2.2 Testing the Stability of Packed Parallel β-Sheets

Two parallel β-sheets each consisting 4 PrP 106-126 fragments are packed against each other and simulated for 1 ns to test the stability of the structure [23]. Simulations are conducted under neural, acidic and alkaline conditions.

### 2.2.3 Elongation of the Fibril

The two edge strands of the packed parallel β-sheets are fixed to represent the formed fibril and a single peptide is put in the same simulation box to test the condition which can accelerate the elongation of the fibril. The simulation time is as long as 10 nm and multiple simulations under different initial conditions are conducted to test the validity of the simulation.

### 2.2.4 Data analysis

To analyze the structure of liquid water, the self diffusion coefficient [30,32], radial distribution function [32], hydrogen bond autocorrelation function [30], hydrogen bond length and angle distribution [32] are used. To analyze the structure of protein, hydrogen bond between protein residues and hydrogen bond between water and protein are analyzed. The root mean square deviation [32] between the protein at a certain time and its initial structure is also used to reveal the behavior of the protein. Since hydrogen bond forming and breaking is a fast process [34], 50 ps simulations were conducted and the structures were saved every 0.02 ps to get such data. Due to the small size of the simulation system, large fluctuations are inevitable. Some information, like the hydrogen bond properties between protein residues can not be extracted from simulation data due to the large noise. Except for the methods mentioned above, there are some other quantities to characterize the structure of water: water molecule network characteristic time [34], the translational order [35] and tetrahedral order [35] which is not used. The structure of protein is analyzed using software "DSSP" [31].

## 3 Results and Discussion

### 3.1 Peptide Conformation

6 ns simulation data is analyzed using software "DSSP"[31]. In both neutral and alkaline conditions, the peptide is quite rigid. In contrast, the peptide is more flexible in acidic condition. In the 6 ns simulation, the peptide adopted extended, helical and hairpin structures. We list the mean number of residues which adopt helical, β and turning structures, respectively (See Table 1). We can also observe a faster increase of the root mean square deviation of the protein with time under acidic condition (See Figure 1).

To see how this happen, we simulated about 10000 water molecules with hydronium ions or hydroxyl ions (pH=2), and analyzed the radial density function (See Figure 2), the self diffusion coefficient and water-water hydrogen bonding properties. The self diffusion coefficient of water increases when hydronium ions or hydroxyl ions are added (See Table 2). Changing in the water radial density function, water-water hydrogen bond autocorrelation, bond length and angle distribution are so small that can not be observed (See Figures 3, 4 and 5).

When simulated with peptide, the water-water hydrogen bond autocorrelation gets stronger (See Figure 5), and the protein-water hydrogen bonding autocorrelation gets weaker under acidic condition (See Figure 6). Considering that the peptide is more flexible under acidic condition, this is consistent with the fact that water plays an important role in maintaining the structure of this peptide which is mainly composed of hydrophobic residues. At the same time, protein-water hydrogen bonds' length and angle distribution get wider under acidic condition (See Figure 7 and 9). However, alkaline condition does not bring obvious change to hydrogen bonding properties. We also calculated the number of the protein-protein and protein-water hydrogen bonds (See Figure 8). The number of protein-water hydrogen bonds is the smallest under acidic condition, alkaline condition also cause the number of protein-water hydrogen bonds to fall, but less obvious.

### 3.2 Stability of the Fibril

The tightly packed β-sheets are very stable because (1) Nearly every possible hydrogen bond between the peptides can form under such condition; (2) The hydrophobic residues can pack together at the center and reduces their direct contact with water. In the three 1 ns simulation, the peptide of the fibril is more stable under alkaline condition than under neutral condition, and it is less stable under acidic condition. However, the fibril structure is not destroyed by adding hydronium ions (See Figure 10). It is possible that there exist a range of pH value within which single peptide is flexible enough and the fibril is stable enough to facilitate the elongation of the fibril.

**3.3 Elongation of the Fibril**

We fix one peptide in each of the two packed β-sheets to represent the edge strands of the fibril, and put a single peptide near (See Figure 11). The structure of the single peptide is obtained by a 4 ns simulation in pure water. Under acidic condition, one of the residues forms hydrogen bond with one of the edge strands of the fibril. Then, inter peptide hydrogen bonds gradually form one by one to the end of the peptide. The single peptide forms β-structure along the edge strands and becomes a new element of the fibril (See Figure 12). This happens in three of the five simulations with different initial conditions. Under alkaline and neutral condition, the hydrogen bonds within this peptide do not break up. The resulting structure under neutral condition is a hairpin but not extended peptide (See Figure 13). Under alkaline condition, the peptide can not become a new element of the structure (See Figure 14).

# 4 Conclusion

We observed the elongation of the PrP 106-126 fragment fibril in molecular dynamics simulation. Basing on this observation and analysis of the simulation data under different conditions, we propose: The hydronium ions changed hydrogen bonding behavior of water and the peptide, making the peptide more flexible thus can be easily influenced by the edge strands of the fibril. When the peptide changed to a β-strand along the edge strands, it becomes a new element of the fibril. This indicates that the fibril propagates by elongation of the fibril at both ends. Only the cross sections of the fibril are infectious. If long fibril has a certain probability (which is proportional to the length) to break into two parts, this fibril elongation mechanism can cause exponential increase of the fibril. Since water molecule network characteristic time [34], the translational order [35] and tetrahedral order [35] have not been used to treat this problem, future work may focused on using these methods to analyze the simulation data to further reveal that how hydronium and hydroxyl ions influence the hydrogen bonding behavior.

# Acknowledgements

This work was supported by the National Natural Science Foundation of China (No. 60025516, No. 10334100) and the Innovation Project of the Chinese Academy of Sciences.

# References

[1] Harris, D. A. (1999) Clin. Microbiol. Rev. 12, 429-444.
[2] Prusiner, S. B. (1991) Science 252, 1515-1522.
[3] Prusiner, S. B. (1998) Proc. Natl. Acad. Sci. USA 95, 13363-13383.
[4] James, T. L., Liu, H., Ulyanov, N. B., Farr-Jones, S., Zhang, H., Donne, D. G., Kaneko, K., Groth, D., Mehlhorn, I., Prusiner, S. B., et al. (1997) Proc. Natl. Acad. Sci. USA 94, 10086-10091.
[5] Riek, R., Hornemann, S., Wider, G., Billeter, M., Glockshuber, R. & Wuthrich, K. (1996) Nature 382, 180-182.
[6] Lopez Garcia, F., Zahn, R., Riek, R. & Wuthrich, K. (2000) Proc. Natl. Acad. Sci. USA 97, 8334-8339.
[7] Zahn, R., Liu, A., Luhrs, T., Riek, R., von Schroet-ter, C., Lopez Garcia, F., Billeter, M., Calzolai, L., Wider, G. & Wuthrich, K. (2000) Proc. Natl. Acad. Sci. USA 97, 145-150.
[8] Calzolai, L. & Zahn, R. (2003) J. Biol. Chem. 278, 35592-35596.
[9] Caughey, B. W., Dong, A., Bhat, K. S., Ernst, D., Hayes, S. F. & Caughey, W. S. (1991) Biochemistry 30, 7672-7680.
[10] Pan, K. M., Baldwin, M., Nguyen, J., Gasset, M., Serban, A., Groth, D., Mehlhorn, I., Huang, Z., Fletterick, R. J., Cohen, F. J. & Prusiner, S. B. (1993) Proc. Natl. Acad. Sci. USA 90, 10962-10966.
[11] Oesch, B., Westaway, D., Walchli, M., McKinley, M. P., Kent, S. B., Aebersold, R., Barry, R. A., Tempst, P., Teplow, D. B., Hood, L. E., et al. (1985) Cell 40, 735-746.
[12] Sunde, M., Serpell, L. C., Bartlam, M., Fraser, P. E., Pepys, M. B. & Blake, C. C. (1997) J. Mol. Biol. 273, 729-739.
[13] Swietnicki, W., Petersen, R., Gambetti, P. & Surewicz, W. K. (1997) J. Bio. Chem. 272, 27517-27520.
[14] Hornemann, S. & Glockshuber, R. (1998) Proc. Natl. Acad. Sci. USA 95, 6010-6014.
[15] DeMarco, M. L. & Daggett, V. (2004) Proc. Natl. Acad. Sci. USA 101, 2293-2298.
[16] Armen, R. S., DeMarco, M. L., Alonso, D. O. V. & Daggett, V. (2004) Proc. Natl. Acad. Sci. USA 101, 11622-11627.
[17] Muramoto, T., Scott, M., Cohen, F. E. & Prusiner, S. B. (1996) Proc. Natl. Acad. Sci. USA 93, 15457-15462.
[18] Salmona, M., Malesani, P., De Gioia, L., Gorla, S., Bruschi, M., Molinari, A., Vedova, F. D., Pedrotti, B., Marrari, M. A., Awan, T., et al. (1999) Biochem. J. 342, 207-214.
[19] Zhang, H., Kaneko, K., Nguyen, J. T., Livshits, T. L., Baldwin, M. A., Cohen, F. E., James, T. L. & Prusiner, S. B. (1995) J. Mol. Biol. 250, 514-526.
[20] Forloni, G., Angeretti, N., Chiesa, R., Monzani, E., Salmona, M., Bugiani, O. & Tagliavini, F. (1993) Nature 362, 543-546.
[21] Tagliavini, F., Prelli, F., Verga, L., Giaccone, G., Sarma, R., Gorevic, P., Ghetti, B., Passerini, F., Ghibaudi, E., Forloni, G., et al. (1993) Proc. Natl. Acad. Sci. USA 90, 9678-9682.
[22] Selvaggini, C., Gioia, L. D., Cantu, L., Ghibaudi, E., Diomede, L., Passerini, F., Forloni, G., Bugiani, O., Tagliavini, F., Salmona, M. (1993) Biochem. Biophys. Re.

Comm. 194. 1380-1386.
[23] Kuwata, K., Matumoto, T., Cheng, H., Nagayama, K., James, T. L. & Roder, H. (2003) Proc. Natl. Acad. Sci. USA 100, 14790-14795.
[24] Petkova, A. T., Ishii, Y., Balbach, J. J., Antzutkin, O. N., Leapman, R. D., Delaglio, F. & Tycko, R. (2002) Proc. Natl. Acad. Sci. USA 99, 16742-16747.
[25] Ma, B. & Nussinov, R. Protein Sci (2002) 11 2335-2350.
[26] Berendsen, H. J. C., van der Spoel, D., van Drunen, R. (1995) Comp. Phys. Commun. 91, 43.
[27] Scott, W.R.P. et al. (1999) J. Phys. Chem. A 103, 3596-3607.
[28] Berendsen, H. J. C., Postma, J. P. M., Gusteren, W. F. and Hermans, J. Intermolecular Forces, edited by B. Pullman (Reidel, Dordrecht, 1981), 331.
[29] HyperChem version 7.0 for Windows, Hypercube Inc., Waterloo, Canada.
[30] Allen, M. P., Tildesley, D. J. (1987) Computer Simulations of Liquids. Oxford: Oxford Science Publications.
[31] Kabsch, W., Sander, C. (1983) Biopolymers 22, 2577-2637.
[32] The self diffusion coefficient can be calculated by mean square displacement using the Einstein relation; The radial distribution function is calculated on the Oxygen atoms; For details about the geometrical criterion to identify a hydrogen bond and the root mean square deviation calculation, please see the gromacs manual at http://www.gromacs.org/documentation/index.php.
[33] van Gunsteren, W. F., Billeter, S. R., Eising, A. A., Hunenberger, P. H., Kruger, P., Mark, A. E., Scott, W. R. P., Tironi, I. G. (1996) Biomolecular Simulation: The GROMOS96 manual and userguide. Zurich, Switzerland: Hochschulverlag AG an der ETH Zurich.
[34] Tarek, M. & Tobias, D. J. (2002) Phys. Rev. Lett. 88 138101.
[35] J. R. Errington & P. G. Debenedetti, (2001) Nature 409, 318-321.

Figure captions:

Figure 1: Root mean square deviation of the protein: The peptide is more flexible under acidic condition and is more rigid under alkaline condition.

Figure 2: RDF of water: The RDF remains unchanged under acidic and alkaline conditions.

Figure 3: Water-water hydrogen bond length distribution: The distribution remains unchanged.

Figure 4: Water-water hydrogen bond angle distribution: The distribution remains unchanged.

Figure 5: Water-water hydrogen bond autocorrelation function: Only with protein and hydronium ions, the hydrogen bond autocorrelation gets stronger.

Figure 6: Protein-water hydrogen bond autocorrelation function: Protein-water hydrogen bond autocorrelation gets weaker under acidic condition.

Figure 7: Protein-water hydrogen bond length distribution: The distribution gets wider under acidic condition.

Figure 8: The number of hydrogen bonds formed between protein and water: The number of protein-water hydrogen bonds decreased under acidic condition.

Figure 9: Protein-water hydrogen bond angle distribution: The distribution gets wider under acidic condition.

Figure 10: The fibril structure: The fibril structure was not destroyed by adding hydronium ions.

Figure 11: The initial structure for simulation: One of the initial structures.

Figure 12: The single peptide becomes a new element of the fibril.

Figure 13: Simulation for fibril elongation under neutral condition: The single peptide formed hydrogen bonds with the edge strands of the fibril under neutral condition, but intra-peptide hydrogen bonds did not break up.

Figure 14: Simulation for fibril elongation under alkaline condition: The single peptide formed a few hydrogen bonds with the edge strands of the fibril under alkaline condition.

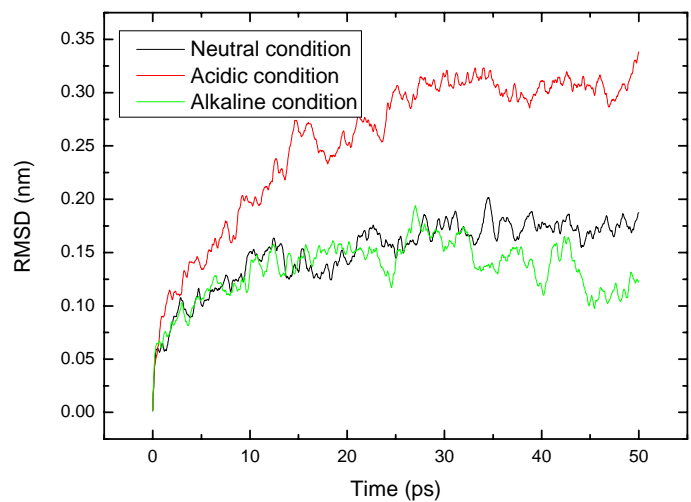

Figure 1

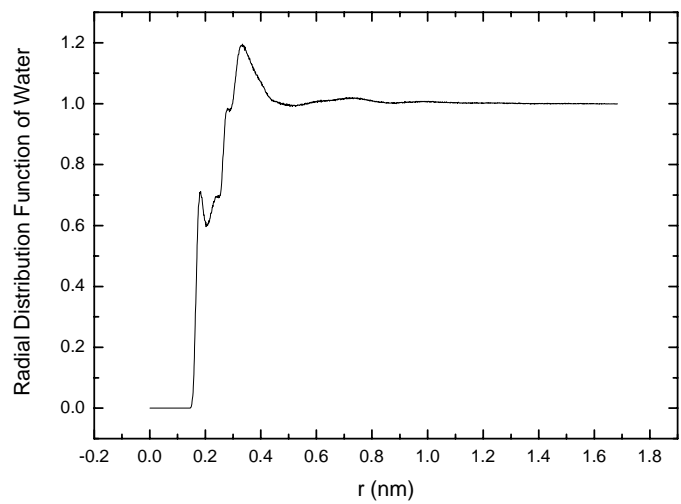

Figure 2

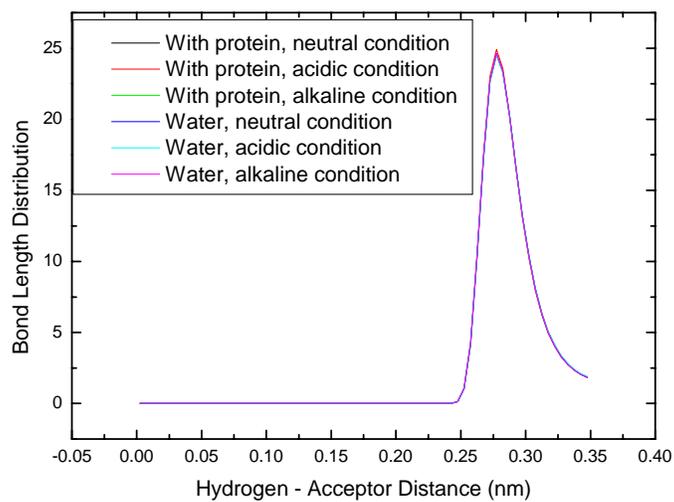

Figure 3

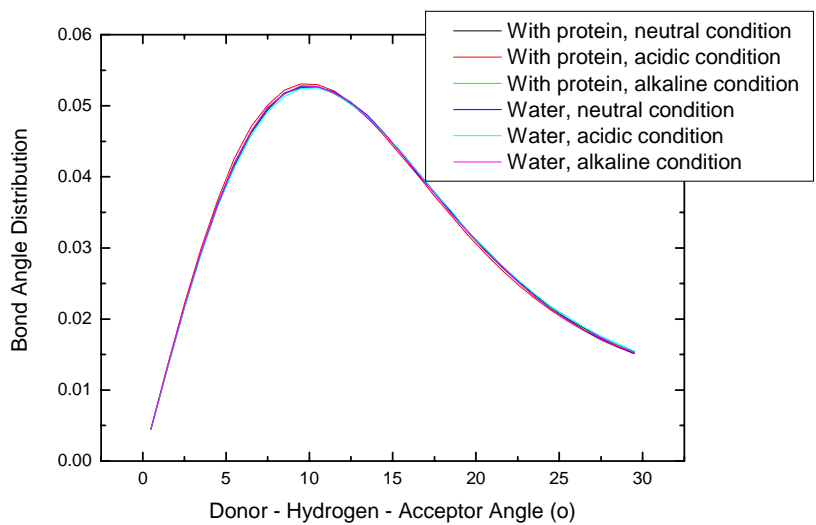

Figure 4

10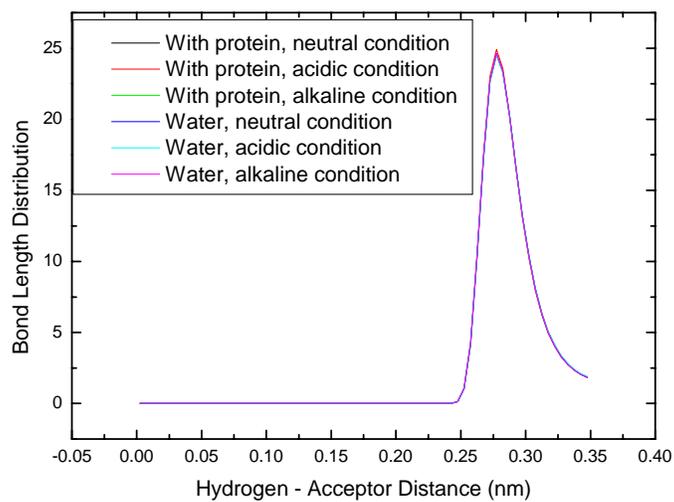

Figure 3

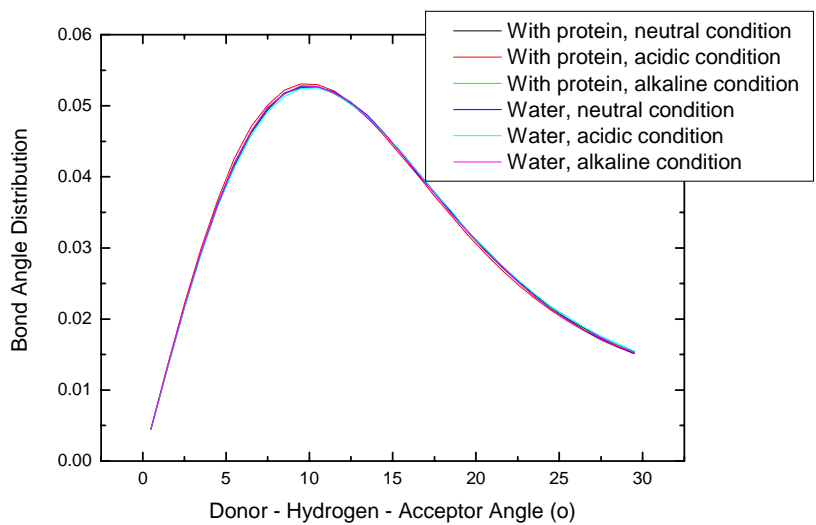

Figure 4

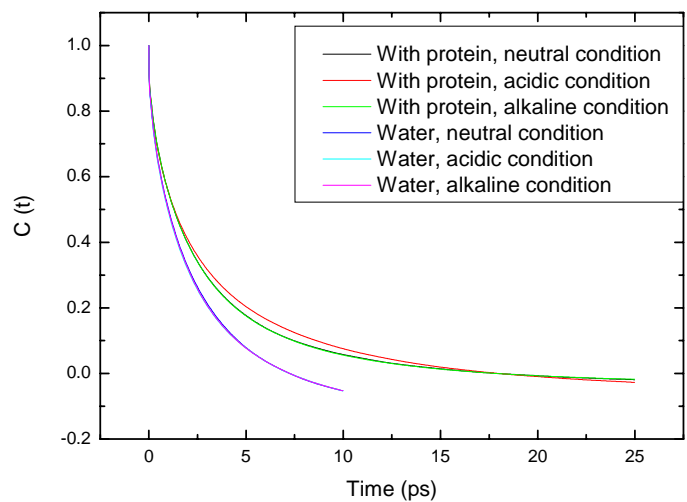

Figure 5

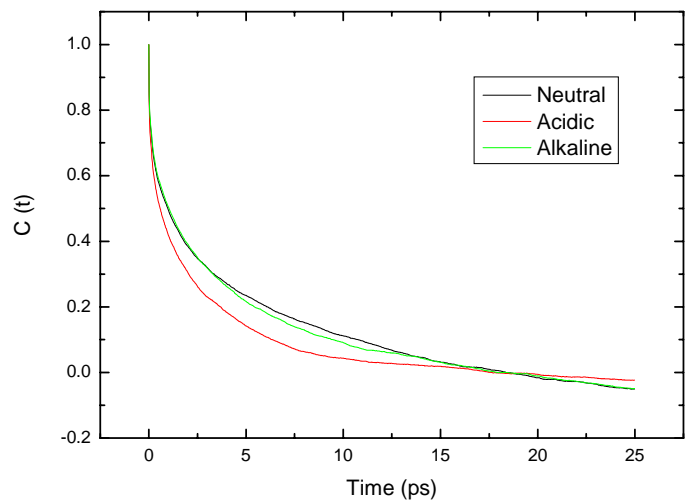

Figure 6

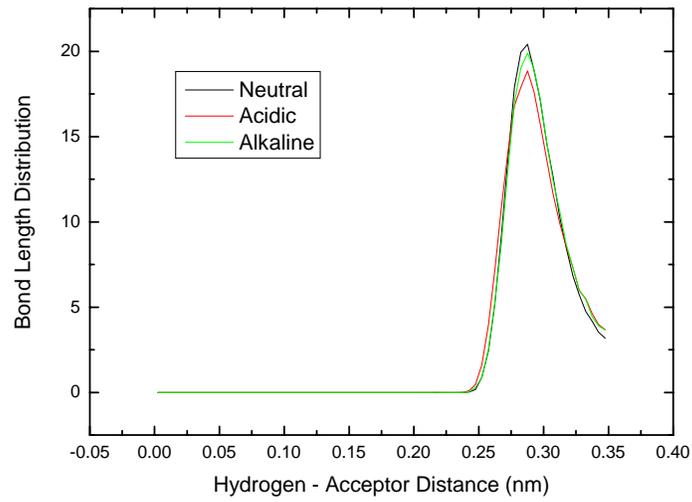

Figure 7

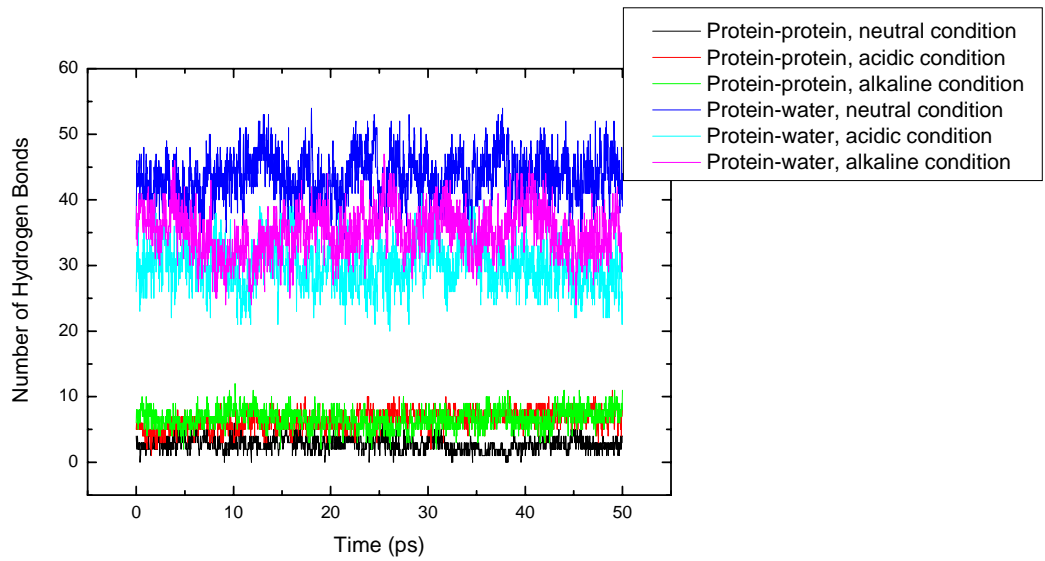

Figure 8

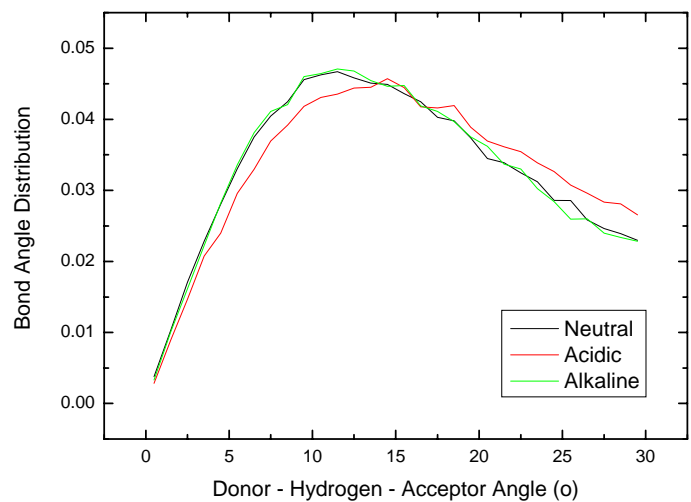

Figure 9

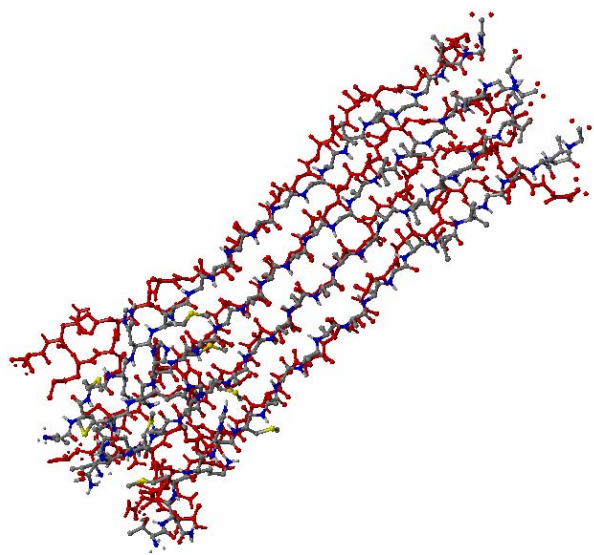

Figure 10

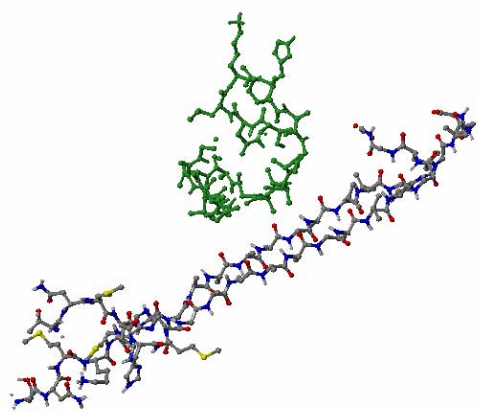

Figure 11

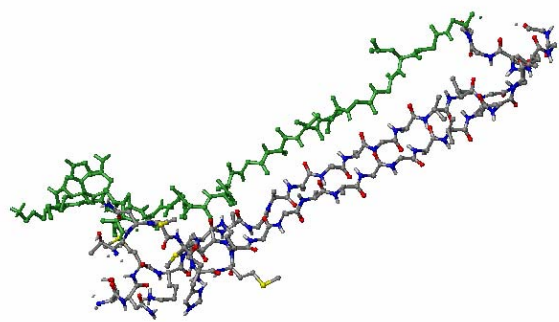

Figure 12

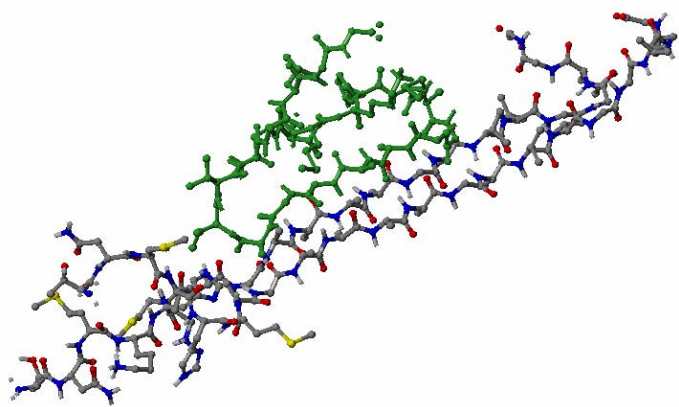

Figure 13

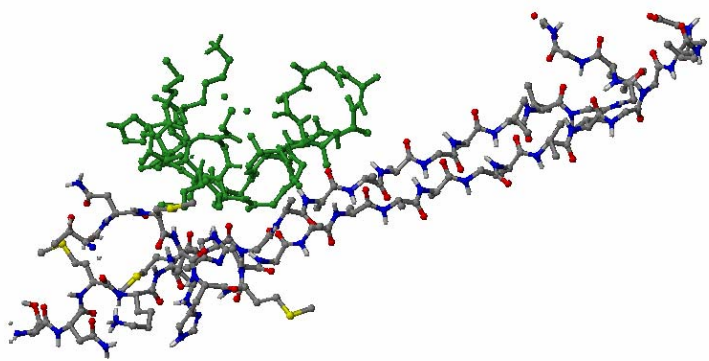

Figure 14

Tables:

Table1: Number of residues which adopt helical, β and turning structures, averaged over 6 ns simulation.

|       | Neutral | Acidic | Alkaline |
|-------|---------|--------|----------|
| Turn  | 1.95    | 2.01   | 1.50     |
| Helix | 0.005   | 0.088  | 0        |
| β     | 2.29    | 2.03   | 2.32     |

Table 2: Water diffusion coefficient under different conditions.

|            | $D\ (cm^2 s^{-1})$            |
|------------|-------------------------------|
| $H_2O$     | $(4.560 \pm 0.082) \times 10^{-5}$ |
| $H_2O/H_3O^+$ | $(4.868 \pm 0.183) \times 10^{-5}$ |
| $H_2O/OH^-$ | $(4.968 \pm 0.015) \times 10^{-5}$ |